\documentclass[aps,pra,twocolumn,superscriptaddress,showpacs]{revtex4-1}
\usepackage{amsmath}
\usepackage{relsize}
\usepackage{amssymb}
\usepackage{amsfonts}
\usepackage{graphicx}
\usepackage[export]{adjustbox}
\usepackage{subfigure}
\usepackage{mathtools}
\newcommand\underrel[3][]{\mathrel{\mathop{#3}\limits_{%
      \ifx c#1\relax\mathclap{#2}\else#2\fi}}}
\DeclareMathOperator{\Li}{Li}
\usepackage{color}

\begin{document}

\title{Universal dynamical scaling of long-range topological superconductors}

\author{Nicol\`o Defenu}
\affiliation{Institut f\"ur Theoretische Physik, Universit\"at 
Heidelberg, D-69120 Heidelberg, Germany}
\author{Giovanna Morigi}
\affiliation{Theoretische Physik, Universit\"at des Saarlandes, D-66123 Saarbr\"ucken, Germany}
\author{Luca Dell'Anna}
\affiliation{Dipartimento di Fisica e Astronomia G. Galilei, Universit\`a degli studi di Padova, via Marzolo 8, 35131 Padova, Italy}
\author{Tilman Enss}
\affiliation{Institut f\"ur Theoretische Physik, Universit\"at 
Heidelberg, D-69120 Heidelberg, Germany}


\begin{abstract}
  We study the out-of-equilibrium dynamics of $p$-wave superconducting quantum wires with long-range interactions, when the chemical potential is linearly ramped across the topological phase transition. We show that the heat produced after the quench scales with the quench rate $\delta$ according to the scaling law $\delta^\theta$, where the exponent $\theta$ depends on the power law exponent of the long-range interactions. We identify the parameter regimes where this scaling can be cast in terms of the universal equilibrium critical exponents and can thus be understood within the Kibble-Zurek framework. When the electron hopping decays more slowly in space than pairing, it dominates the equilibrium scaling. Surprisingly, in this regime the dynamical critical behaviour arises only from paring and, thus, exhibits anomalous dynamical universality unrelated to equilibrium scaling. The discrepancy from the expected Kibble-Zurek scenario can be traced back to the presence of multiple universal terms in the equilibrium scaling functions of long-range interacting systems close to a second order critical point.
\end{abstract}
\date{\today}

\maketitle
One of the major challenges of contemporary physics is the identification of quantum phases of matter, which can serve as platforms for quantum computers. In this perspective, topological superconductors\,\cite{Hasan2010,Bernevig2013} are promising constituents for quantum devices\,\cite{Nayak2008,Terhal2015,Kraus2013a,Mazza2013}, thanks to the presence of gapless Majorana modes, the so-called Majorana zero modes (MZM), which are localised at the chain edges and topologically protected.  Since the first theoretical evidence of MZMs in superconducting wires\,\cite{Kitaev2001}, several experimental platforms have revealed consistent signatures of Majorana physics both in one-dimensional\,\cite{Mourik2012,Deng2012,Das2012,Albrecht2016} and two-dimensional\,\cite{Wang2012,He2014,Sun2016,He2017} geometries. More recently, models of $p$-wave superconducting wires with long-range (LR) deformations have shown more robust topological properties\,\cite{Viyuela2015,Viyuela:2018fpv}, while strong enough LR pairing effects alter the nature of the topological phase\,\cite{Vodola2014,Lepori2015,Lepori2017,Lepori2017add, Alecce2017} and the spreading of correlations\,\cite{Foss-Feig2015,Cevolani2015,Vodola2016}. 
Experimental realisations of LR topological superconductors employ one-dimensional arrays of magnetic impurities on top of a conventional superconducting substrate\,\cite{Nadj-Perge2014,Pawlak2016,Ruby2017}, leading to the realisation of an effective Kitaev Hamiltonian with both LR pairing and LR hopping\,\cite{Pientka2013,Klinovaja2013,Pientka2014,Neupert2016}. In this context, understanding slow variations (quenches) of control fields in quantum systems is fundamental to adiabatic protocols \,\cite{Nielsen2000}, since Majorana excitations cannot be realised by sudden manipulations of the system\,\cite{Perfetto2013}. These investigations constitute a fundamental contribution towards the understanding of dynamical scaling for quenches across topological phase transitions\,\cite{Ueda2010}. 

\begin{figure}
\includegraphics[scale=.3]{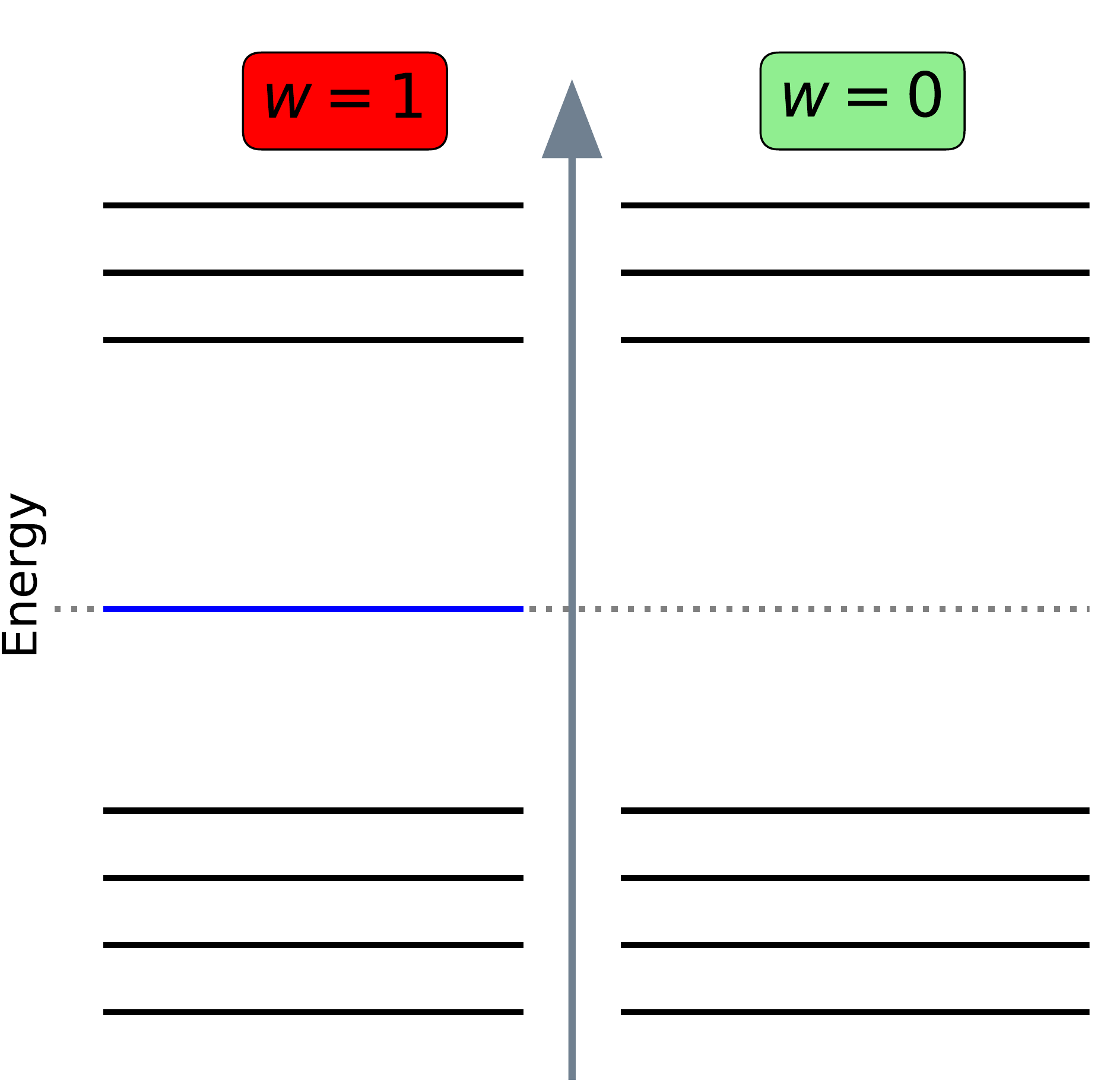}
\caption{\label{Fig1}(color online) (a) The energy spectrum of a one-dimensional topological superconductor for chemical potential $|\mu|<\mu_{c}$ with bulk topological invariant $w=1$. The blue solid line represents the degenerate groundstate, which hosts MZMs. (b) For $|\mu|>\mu_{c}$ there is a single ground state with $w=0$.}
\end{figure}

In this Letter we characterise the out-of-equilibrium dynamics of a $p$-wave superconducting quantum wire with long-range interactions, whose chemical potential is linearly ramped across the equilibrium critical point. We determine the density of defects produced after the ramp and show that it scales as a power of the quench rate. We then connect the power-law exponent with the equilibrium critical properties and topological features and determine the phase diagram for the dynamics as a function of the decay exponent of the hopping and pairing terms. 

\begin{figure}[t]
\includegraphics[width=.35\textwidth]{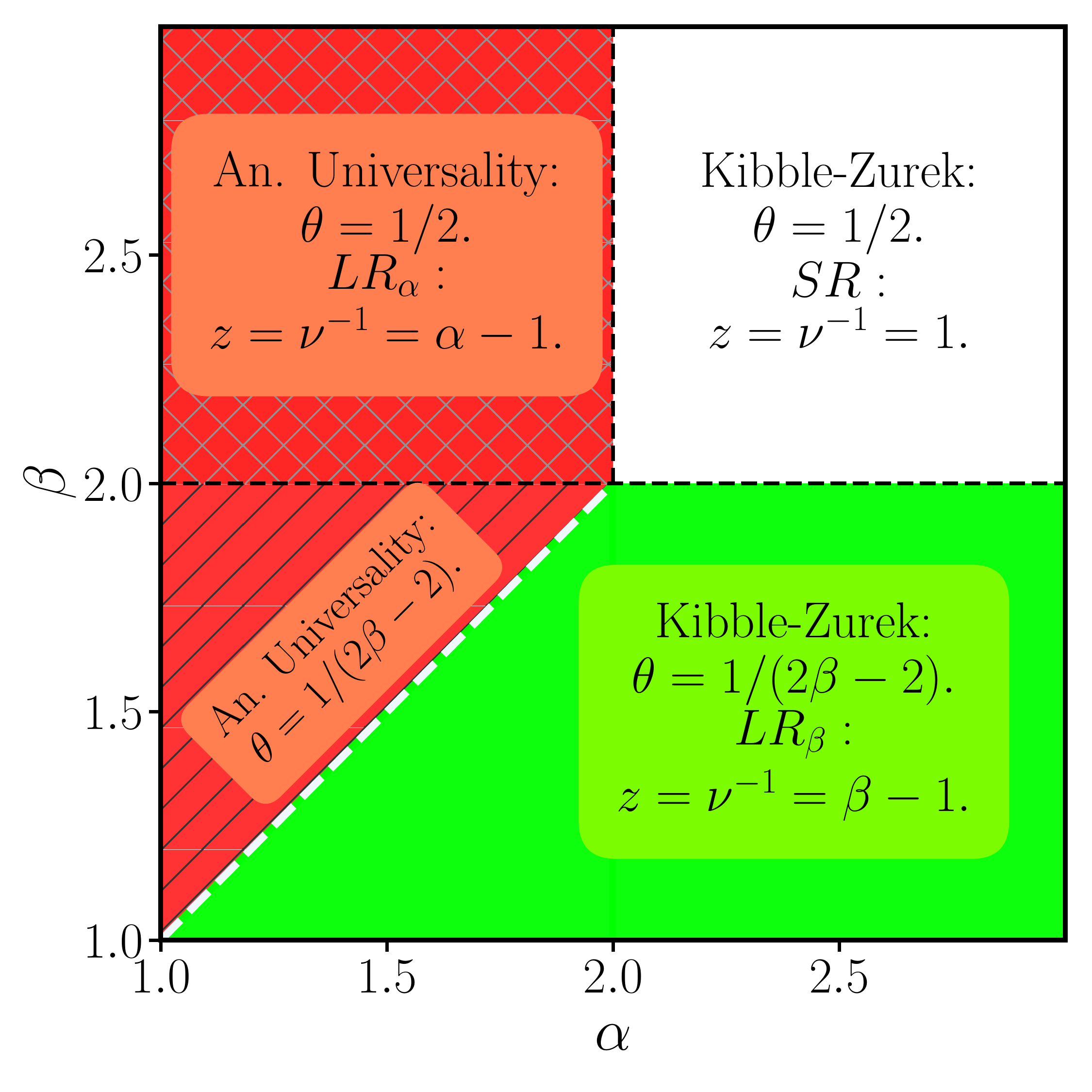}
\caption{\label{Fig1b} Phase diagram of the $p$-wave superconducting Hamiltonian \eqref{h_klr} as a function of the exponents of the power-law decay of hopping, $\alpha$, and pairing, $\beta$. The labels indicate (i) short-range (SR) universality, (ii) hopping dominated long-range universality (LR$_{\alpha}$) for $\alpha<\beta$, (iii) pairing dominated long-range universality (LR$_{\beta}$) for $\beta<\alpha$.  The scaling of defect generation is described by the exponent $\theta$, see Eq.~\eqref{exc_scaling}. In the green region $\beta<\min(\alpha,2)$, dominant pairing determines both equilibrium exponents and standard Kibble-Zurek dynamical scaling. In the red region $\alpha<\min(\beta,2)$, instead, dominant hopping determines equilibrium exponents but dynamics is governed by LR pairing, giving rise to anomalous universal scaling (an. universality) $\theta=1/2$ for $\beta>2$.}
\end{figure}

We consider spinless electrons hopping across the $N$ sites of a linear chain in presence of $p$-wave pairing. The Hamiltonian reads
\begin{align}
\label{h_klr}
\hat H=-\sum_{i}\Bigl[\sum_{r>0}\left(j_{r}\hat c^{\dagger}_{i}\hat c_{i+r}+\Delta_{r}\hat c^{\dagger}_{i}\hat c^{\dagger}_{i+r}+{\rm H.c.}\right)+\mu\hat c^{\dagger}_{i}\hat c_{i}\Bigr]+\mathcal C,
\end{align}
where operators $c^{\dagger}_{i}$ create a fermion at site $i$ and fulfill the anticommutation relations $\{c_i,c_j^\dagger\}=\delta_{ij}$. Here, $\mu$ denotes the chemical potential, $\mathcal C=N\mu/2$ is an energy offset, 
$j_{r}$ and $\Delta_{r}$ are the hopping and pairing amplitudes, respectively, and depend on the intersite distance $r$ according to the power laws (reported here for open boundary conditions):
\begin{align}
j_{r}^{\alpha}&=\frac{J}{N_{\alpha}}\,\frac{1}{r^{\alpha}},&\Delta_{r}^{\beta}&=\frac{d}{N_{\beta}}\,\frac{1}{r^{\beta}}\,,
\label{dr}
\end{align}
with the hopping exponent $\alpha>1$, the pairing exponent $\beta>1$, the coefficients $J,d>0$, and $N_\gamma=2\sum_{r=1}^{N/2} r^{-\gamma}$ the Kac scaling, which guarantees extensivity of the energy\,\cite{Campa2014}. For sufficiently fast decaying interaction and hopping terms the system possesses two different phases separated by the quantum critical point $\mu_c=2J$\,\cite{Kitaev2001}.  In the thermodynamic limit the two topological phases can be distinguished by the bulk topological invariant $w$: For $|\mu|>\mu_{c}$ the ground state is nondegenerate and $w=0$; in the nontrivial phase $|\mu|<\mu_{c}$ the bulk topological invariant $w=1$, and the ground state is doubly degenerate and can host MZMs, see Fig.\,\ref{Fig1}. At finite size $N$ the spectrum is always gapped and for open boundary conditions the MZMs remain localized at the edges of the chain. The presence of the LR pairing and hopping terms in Eq.\,\eqref{h_klr} does not alter this phase diagram nor the values of the bulk topological invariant as long as $\alpha,\beta>1$\,\cite{Vodola2014,Viyuela:2018fpv,Alecce2017}.  Nonetheless, LR connectivity modifies the universal critical behaviour of the model by changing the critical exponents. The resulting phases are displayed in Fig.~\ref{Fig1b}. Note that the equilibrium phase diagram of the long-range Kitaev chain radically differs from the one of the long-range quantum Ising model\,\cite{Defenu2016,Defenu2017a}. 

In the following, we analyse the dynamics during slow variations of the chemical potential across the critical value according to 
\begin{align}
\label{ramp}
\mu=\mu_{c}-\delta\cdot t\,,
\end{align}
where time varies in the interval $[-\mu_c/\delta,\mu_c/\delta]$, i.e., from the topologically trivial phase $\mu=2\mu_c$ deep into the nontrivial phase at $\mu=0$. We note that the time-dependent dynamics have been solved for an Ising model in transverse field\,\cite{Kolodrubetz2012}, which can be mapped to the Kitaev model for $\alpha,\beta\to\infty$\,\cite{Fradkin1989}. Below we derive an exact solution which is valid for general $\alpha,\beta >1$ and in the thermodynamic limit. This solution allows us to determine the thermodynamic functions after the ramp. For this purpose we rewrite the Hamiltonian \eqref{h_klr} using momentum-space operators $\hat c_{k}=e^{i\pi/4}\sum_{r\in\mathbb{Z}}c_{r}e^{ikr}/\sqrt{N}$ with $k\in[-\pi,\pi)$. Using the spinor representation $\hat\psi_k= (\hat c_{k},\hat c_{-k}^{\dagger})^T$, the Hamiltonian is the sum of $2\times 2$ block matrices $\mathcal H_k\equiv \boldsymbol{h}_{k}(t)\cdot\boldsymbol{\hat\sigma}$,
\begin{align}
\label{h_klr_compact}
\hat H(t)=\sum_{k} \hat\psi_k^\dagger\hat{\mathcal H}_k(t)\hat\psi_k \,,
\end{align}
where $\boldsymbol{\hat\sigma}$ is the vector of the Pauli matrices $\hat\sigma^{j=1,2,3}$ and $\boldsymbol{h}_{k}(t)=(\Delta_{\beta}(k),0,\varepsilon_{\alpha}(k,t))$ is the pseudo-spin vector.
Its elements depend on $\varepsilon_{\alpha}(k,t)=\mu(t)/2-j_{\alpha}(k)$ and on the momentum-space hopping and pairing coefficients:
\begin{align}
j_{\alpha}(k)&=J\,\text{Re}\left[\Li_{\alpha}(e^{ik})\right]/\zeta(\alpha)\,,\label{funcj}\\
\Delta_{\beta}(k)&=d\,\text{Im}\left[\Li_{\beta}(e^{ik})\right]/\zeta(\beta)\,,\label{funcd}
\end{align}
where $\Li_\alpha(z)$ denotes the polylogarithm and $\zeta(\alpha)$ the Riemann zeta function\,\cite{Abramowitz1964}. The Hamiltonian is diagonalized in terms of the fermionic quasiparticle operators $\hat \gamma_k(t)=u_k(t)\hat c_k+v_{-k}^*(t)\hat c_{-k}^\dagger$ to obtain $\hat H=\sum_{k}\omega_{k}(t)\bigl(\hat \gamma^{\dagger}_{k}(t)\hat\gamma_{k}(t)-\frac{1}{2}\bigr)$ with the quasiparticle spectrum $\omega_{k}(t)=2\sqrt{\varepsilon_{\alpha}(k,t)^{2}+\Delta_{\beta}(k)^{2}}$. The pseudo-spin vector $\boldsymbol{h}_{k}(t)$ identifies a direction in the two-dimensional plane of the Hamiltonian space. At a given instant of time, integrating the angle $\theta_{k}={\rm atan}(h_k^1/h^3_k)={\rm atan}(\Delta_\beta(k)/\varepsilon_\alpha(k))$ over the Brillouin zone yields the bulk topological invariant $w=\oint d\theta_{k}/(2\pi)$ of the corresponding equilibrium phase. 

\begin{figure*}[t]
\subfigure[
]{\includegraphics[scale=.23]{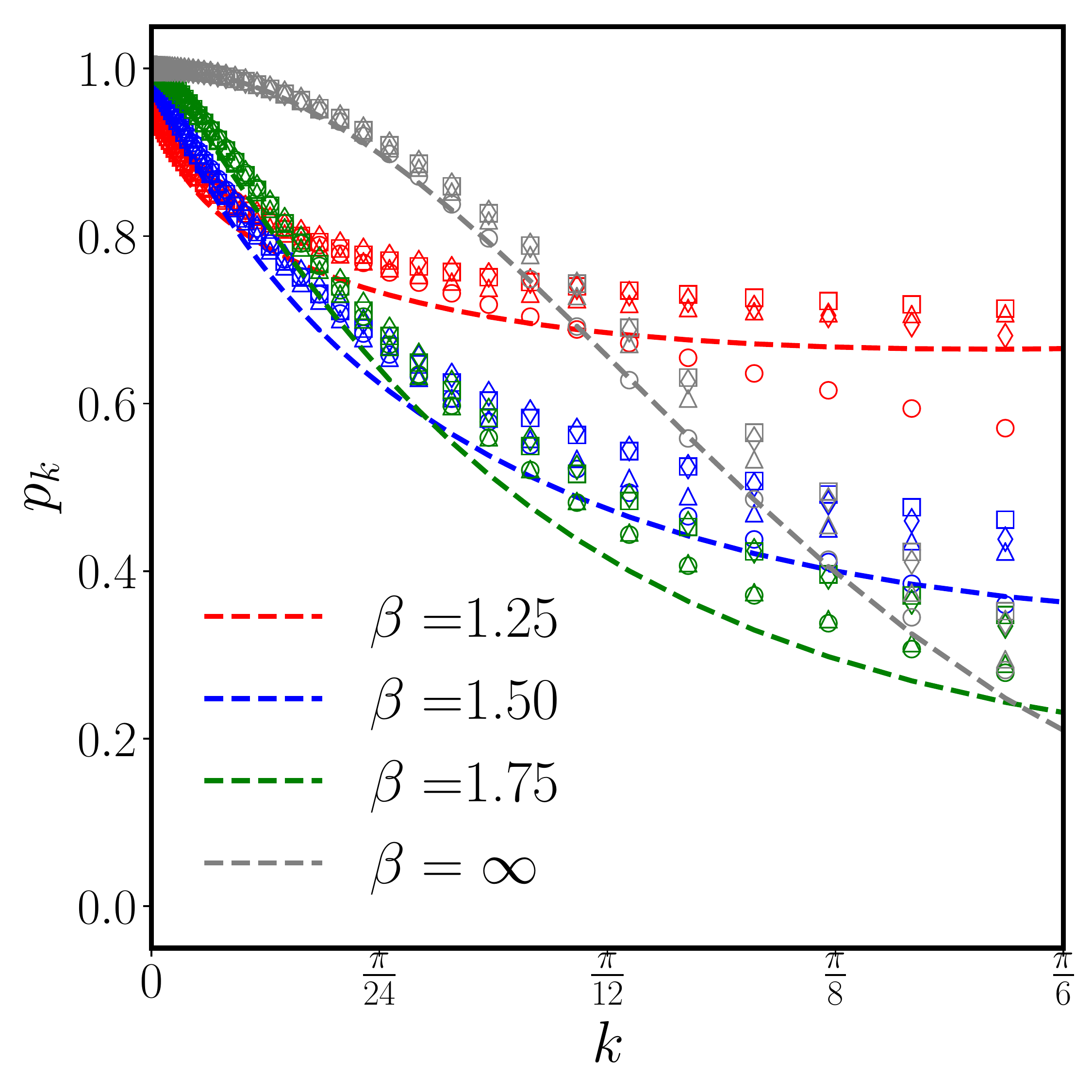}\label{Fig2a}}
\subfigure[
]{\includegraphics[scale=.23]{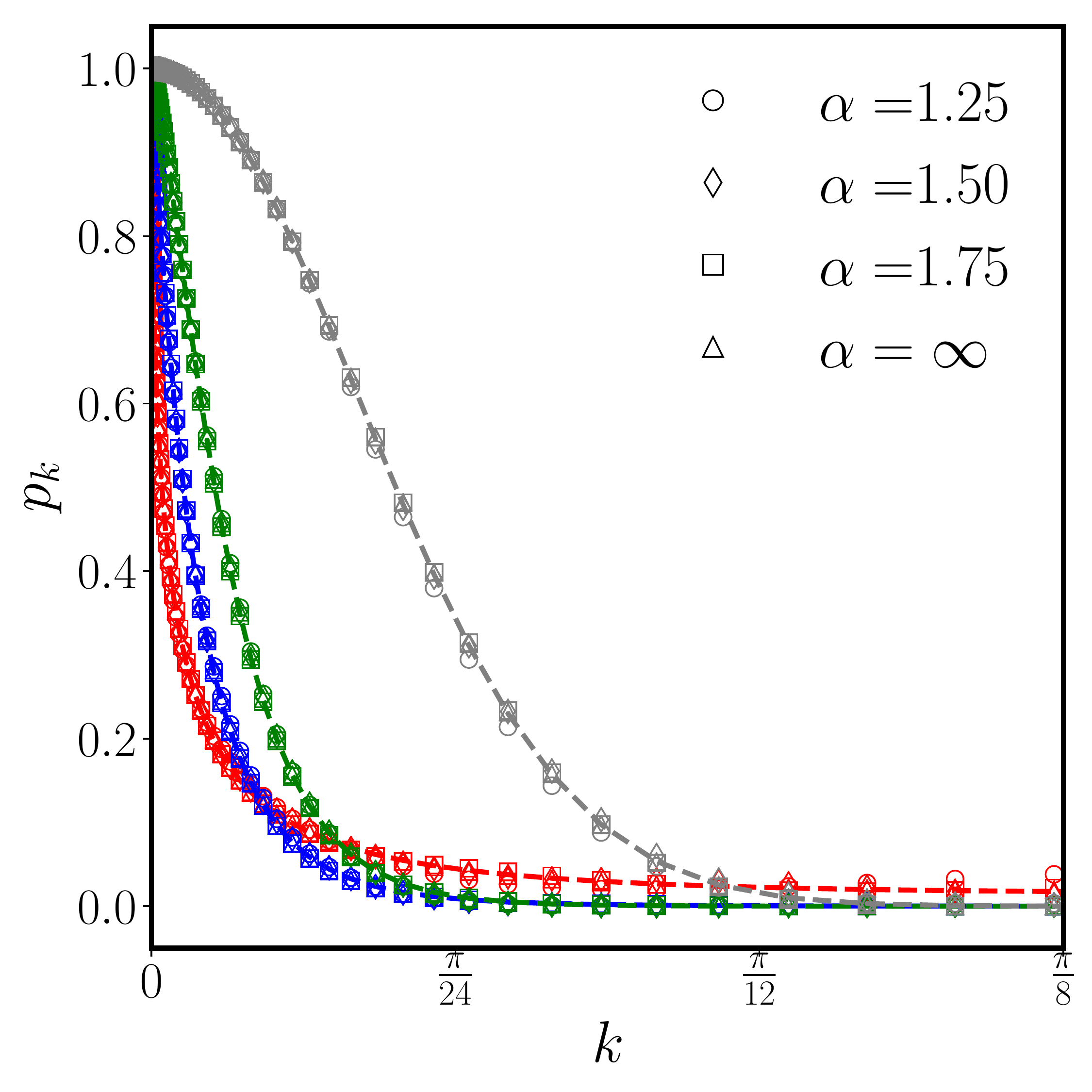}\label{Fig2b}}
\subfigure[
]{\includegraphics[scale=.23]{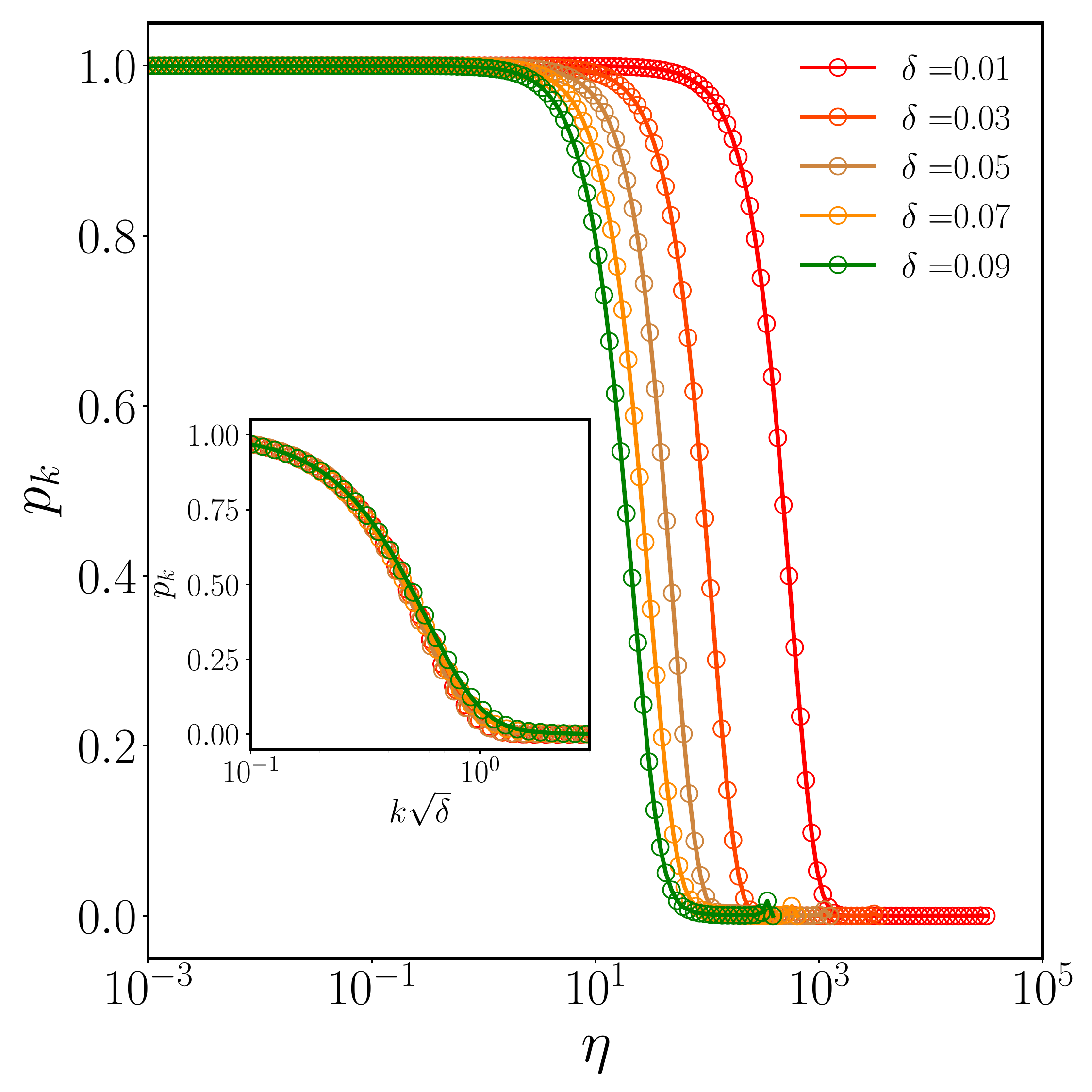}\label{Fig2c}}
\caption{\label{fig2}. The excitation probability $p_{k}$ of the Bogolyubov modes after a ramp of the chemical potential, see Eq.\,\eqref{ramp},  obtained by the numerical solution of Eq.\,\eqref{dyn_sys} for intermediate and small ramp velocities $\delta=0.5,0.05$ in panels (a) and (b) respectively. For each curve the value of $\alpha=(1.25,1.5,1.75,\infty)$ is indicated by round, diamond, square and triangular symbols, see the legend of panel (b), while the value of $\beta$ is given by the color: red, blue, green and gray for, respectively, $\beta=(1.25,1.5,1.75,\infty)$. Away from the adiabatic limit, panel (a), the numerical results (symbols) do not collapse on the theoretical expectations given by Eq.\,\eqref{exc_prob} (dashed lines). Even so, the predicted trend for $k_{th}$ is clearly visible with the lower $\beta$ values becoming increasingly athermal. For the very slow quench of panel (b) the theoretical prediction in Eq.\,\eqref{exc_prob} is almost perfectly obeyed, with no correction arising even at small $\alpha$s. Panel (c) exemplifies the $\alpha<\beta$ case, where anomalous dynamical scaling exists, for several $\delta$ values with $\alpha=1.25$ and $\beta=\infty$. The excitation probability $p_{k}$ is reported as a function of the universal variable $\eta$, obtained by equilibrium scaling, and no collapse is found. Conversely, the curves collapse perfectly on each other, when plotted as function of the universal variable $k/\sqrt{\delta}$, obtained with the proper dynamical scaling $z_{d}=\nu_{d}^{-1}=1$. These results prove the existence of a universal dynamical scaling different from the standard Kibble-Zurek scenario.}
\label{Fig2}
\end{figure*}

The dynamics of the Kitaev chain can be exactly described by the Heisenberg equations of motion for the original creation and annihilation operators, ${\rm i}d\hat {c}_{k}/d t=[\hat c_{k},\hat H]$. These equations can be cast into a matrix evolution for the Bogolyubov coefficients,
\begin{align}
\label{dyn_sys}
i\frac{d}{dt}\begin{pmatrix}u_{k}\\v_{k}\end{pmatrix}=\begin{pmatrix}\varepsilon_{\alpha}(k,t) & -\Delta_{\beta}(k)\\
\Delta_{\beta}(k) & \varepsilon_{\alpha}(k,t)
\end{pmatrix}\begin{pmatrix}u_{k}\\v_{k}\end{pmatrix},
\end{align}
which can be mapped into the Landau-Zener form\,\cite{dziarmaga2005,dziarmaga2010,Dutta2017}, see App.\,\ref{lz}. 
The excitation probability $p_k(t)$  can be computed exactly\,\cite{Damski2005a,Bialonczyk:2018sbn}, see App.\,\ref{defect_sc}. For a slow quench to the final time $\tau_{f}=\mu_c/\delta$, the excitation probability is well approximated by the Landau-Zener formula 
\begin{align}
\label{exc_prob}
p_{k}\simeq\exp\left(-\frac{\pi \Delta_{\beta}(k)^{2}}{\delta}\right)\,,
\end{align}
which becomes exact for $k\ll\frac{\pi}{2}$ in the slow ramp limit $\delta\to0$.   From Eq.\,\eqref{exc_prob} we find population inversion $p_k\ge 1/2$ when $|k|<k_{\rm th}$. The threshold value $k_{\rm th}$ is determined analytically from a low-momentum expansion and reads $k_{\rm th}=[\delta\log(2)/(\pi c)]^{\theta}$ with
\begin{align}
c=\begin{cases}\left(\frac{\cos(\beta\pi/2)\Gamma(1-\beta)}{\zeta(\beta)}\right)^{2} & \text{for}\,\,1<\beta<2,\\
\left(\frac{\zeta(\beta-1)}{\zeta(\beta)}\right)^{2} & \text{for}\,\,\beta>2.
\end{cases}
\end{align}
It is remarkable that  $k_{\mathrm{th}}\to\infty$ for $\beta\to 1^+$, corresponding to negative temperatures. In the small $\delta$ limit, this effect is only visible very close to the singular limit $\beta\gtrsim1$, while for intermediate $\beta$'s the tendency is reversed, see Fig.\,\ref{Fig2}. These results have been numerically verified taking the full $k$ dependence of $\Delta_\beta(k)$ into account. We note that stable athermal distributions are generally expected in systems with diverging long-range interactions\,\cite{Kastner2011}, and the case we analyse here seems to be no exception. 

Using Eq.~\eqref{exc_prob} we can derive several thermodynamic properties for asymptotically slow drive $\delta\to0$. We discuss here the excitation density\,\cite{zurek2005,chandran2012,dziarmaga2010,polkovnikov2005},
\begin{align}
  n_{\mathrm{exc}}=\frac1N \sum_{k}\langle\gamma_{k}^{\dagger}\gamma_{k}\rangle=\frac1N\sum_{k}p_{k}\,,
  \label{eq:nt}
\end{align}
which we compute in the thermodynamic limit, thus replacing the sum by an integral over the interval $k\in[-\pi,\pi)$. In the $\delta\to 0$ limit the exponent of $p_k$ in Eq.\,\eqref{exc_prob} diverges and the total contribution to the integral only comes from the saddle point. Expanding around the vanishing effective frequency, we find the scaling law
\begin{align}
  \label{exc_scaling}
\lim_{\delta\to 0}n_{\mathrm{exc}}\sim \delta^\theta\quad\mathrm{with}\quad
  \theta=
  \begin{cases}
    (2\beta-2)^{-1}& \text{for }\beta\leq 2, \\
    1/2& \text{for}\beta>2.
  \end{cases}
\end{align}
At the border $\beta=2$ we find $n_{\rm ex}\propto \sqrt{\delta}/\log\delta$. These scalings are valid irrespectively of the value of $\alpha$, since the defect density solely depends on $\beta$. The corresponding dynamical phase diagram is depicted in Fig.~\ref{Fig1b}. Remarkably, the dynamical phases for $\alpha>2$ correspond to the regions of the equilibrium phase diagram, but in the hopping dominated regime $\alpha<2$ a different universal dynamical scaling arises. Such universal dynamical scaling with $\beta$ cannot be related to the equilibrium critical exponents, which involve $\alpha$, as generally happens in Fermi systems\,\cite{dziarmaga2010,polkovnikov2005,de_grandi2010} and its appearance can be traced back to the violation of the equilibrium scaling hypotheses due to the LR nature of the interactions. 

In order to verify our analytical prediction, we numerically integrated Eq.\,\eqref{dyn_sys}. Initially at $t=t_{i}=-\mu_{c}/\delta$ the system is at equilibrium with Bogolyubov coefficients
$u^{i}_{k},v^{i}_{k}=\cos\frac{\theta_{i}(k)}{2},
\sin\frac{\theta_{i}(k)}{2}$
where $\tan\theta_{i}(k)=\frac{\Delta_{\beta}(k)}{\varepsilon_{\alpha}(k)}$ and $\mu(t_{i})=2\mu_{c}=4J$. The Bogolyubov coefficients are then evolved numerically according to Eq.\,\eqref{dyn_sys} for a grid of $k$ points in the interval $[-\pi,\pi)$. Due to non-adiabatic effects arising during the critical stage of the dynamics $t\simeq0$, the resulting amplitudes at the final time $t_{f}=\mu_{c}/\delta$ differ from the ones of the equilibrium Hamiltonian with $\mu(t_{f})=0$. In order to quantify these deviations, we consider the excitation probability of each state $k$, after the slow ramp, with respect to its equilibrium ground state,
\begin{align}
\label{expl_exc_prob}
p_{k}=1-|u^{f}_{k}u^{*}_{k}(t_{f})+v^{f}_{k}v^{*}_{k}(t_{f})|^{2},
\end{align}
where $(u^{f}_{k},v^{f}_{k})$ are the equilibrium Bogolyubov amplitudes at $\mu=0$, while $(u_{k}(t_{f}),v_{k}(t_{f}))$ are the ones at the end of the dynamical evolution.

The excitation probability at the end of the slow quench, calculated according to Eq.\,\eqref{expl_exc_prob}, is shown in Fig.\,\ref{Fig2} as a function of the momentum $k$ for $\delta=0.5,\,0.05$ in panels (a) and (b), respectively. As the dynamical protocol crosses the $\mu_{c}=2J$ critical point, only low momentum modes $k\approx 0$ become soft during the dynamics. Indeed, Eq.\,\eqref{exc_prob} only applies to excitations modes with $k<\pi/2$, as follows from the Landau-Zener mapping, see App.\,\ref{lz}, while high energy modes $k>\pi/2$ remain adiabatic and their excitation probability is not shown. Numerical points for the Bogolyubov modes excitation probability for $\alpha=(\infty,1.75,1.50, 1.25)$ are shown by squares, crosses, circles and triangles respectively (see legend of panel b), while the different values of $\beta=(\infty, 1.75,1.5,1.25)$ are reported respectively from top to bottom (gray, green, blue and red). In Fig.\,\ref{Fig2b} we observe almost perfect agreement with the predictions of Eq.\,\eqref{exc_prob} in the slow ramp case $\delta=0.05$.  Indeed, corrections from finite and slightly asymmetric endpoints $t_{i}$ and $t_{f}$ do not influence the universal behaviour obtained in the $\delta\to 0$ limit at small momenta and can be safely discarded, see App.\,\ref{fr_corr}.

The result of Eq.\,\eqref{exc_scaling} contradicts the result found using adiabatic perturbation theory, which produces the Kibble-Zurek relation between the universal slow dynamics and the equilibrium critical exponents $\theta=\nu/(1+z\nu)$\,\cite{dziarmaga2010,polkovnikov2005,de_grandi2010}. Here, the critical exponents $z\nu$ and $z$ describe the scaling of the spectrum at the critical point, $\omega_{k=0}\propto |\mu-\mu_{c}|^{z\nu}$ and  $\omega_{k\to 0}\propto k^{z}$. In particular, at lowest order in the adiabatic expansion, the excitation probability $p_{k}=|\alpha_{k}|^{2}$ of the Bogolyubov quasi-particle states $|k\rangle=\hat{\gamma}_{k}^{\dagger}|0\rangle$ are given by the squared transition amplitudes induced by the perturbation operator $\hat{\partial}_{\mu}=\partial \hat{H}(\mu)/\partial\mu$ over the Bogolyubov vacuum $|0\rangle$ integrated over the whole dynamical trajectory
 \begin{align}
 \label{ad_dens_exc}
\alpha_{k}\approx \int\langle k|\hat{\partial}_{\mu}|0\rangle e^{\frac{i}{\delta}\int^{\mu}\left(E_{k}(\mu')-E_{0}(\mu')\right)d\mu'}d\mu,
\end{align} 
where $E_{k}(\mu)$ is the energy of the state $|k\rangle$. In the $\delta\to 0$ limit, the saddle-point approximation holds and the integral only receives contribution from the vanishing gap region of the trajectory, i.e. the critical point. There, one can employ the universal scaling relations\,\cite{dziarmaga2010,polkovnikov2005,de_grandi2010}
\begin{align}
\label{scaling_forms}
E_{k}(\mu)-E_{0}(\mu)&=\omega_{k}\approx \Delta\,F\left(\Delta/k^{z}\right)\\
\label{scaling_forms2}
\langle k|\hat{\partial}_{\mu}|0\rangle&\approx\frac{\Delta}{|\mu-\mu_{c}|k^{z}}G\left(\Delta/k^{z}\right)
\end{align}
 where $\Delta$ is the minimal gap $\Delta\propto |\mu-\mu_{c}|^{z\nu}$. Inserting Eqs.\,\eqref{scaling_forms}\,and \eqref{scaling_forms2} into Eq.\,\eqref{ad_dens_exc} and making the integration dimensionless, one finds the universal scaling variables $\eta=k\delta^{-\frac{\nu}{1+z\nu}}$ and $\zeta=k^{1/\nu}(\mu-\mu_{c})$. Rephrasing the adiabatic perturbation theory expression for the defect density $n_{\mathrm{exc}}\approx\int dk |\alpha_{k}|^{2}/(2\pi)$ in terms of the universal variables $\eta$ and $\zeta$ immediately leads to the Kibble-Zurek result  $\theta=\nu/(1+z\nu)$, see Eq.\,\eqref{ad_dens_exc} and Refs.\,\cite{dziarmaga2010,polkovnikov2005,de_grandi2010}. Since for the p-wave superconducting Hamiltonian in Eq.\,\eqref{h_klr} one has $z\nu=1$ and $z=\phi-1$, where $\phi=\mathrm{min}(\alpha,\beta)$, we can conclude that the scaling exponent $\theta$ in Eq.\,\eqref{exc_scaling} is inconsistent with the Kibble-Zurek scaling in the region $\alpha<\beta$.
 
We refer to this unexpected behaviour as \emph{anomalous universal scaling}. We report its extent in Fig.\,\ref{Fig2c} where the example case of LR hopping $\alpha=1.25$ and short range pairing $\beta=\infty$, well inside the anomalous universal scaling region, is studied. The excitation probability is reported as a function of the universal scaling variable $\eta=k\delta^{-\frac{\nu}{1+z\nu}}$ for several $\delta$ values. Remarkably, for this scaling the curves do not collapse, see Fig.\,\ref{Fig2c}. Instead, universality is recovered when one considers the proper dynamical exponent $z_{d}=\nu_{d}^{-1}=1$, for nearest neighbour pairing which is the only responsible for the dynamics Fig.\,\ref{Fig2c}. Indeed, perfect collapse of the excitation probabilities for various $\delta$ is observed in terms of the correct scaling variable $k/\delta^{\nu_{d}/(1+z_{d}\nu_{d})}=k/\delta^{1/2}$, see the inset in Fig.\,\ref{Fig2c}. 
 
In conclusion, we have demonstrated that long-range coupling terms can lead to a novel scaling behaviour of heat produced by slow quenches in critical topological superconductors. We have shown that this behaviour cannot be understood within the framework of the Kibble-Zurek scaling. In particular, the introduction of long-range hopping terms which decay slower than the pairing couplings modifies the equilibrium critical properties but not the dynamical critical exponent $\theta$. In the traditional case, for dominant pairing $\alpha>\beta$, the MZMs are gapped in the broken phase and, approaching the critical point, the gap closes and couples them into a single Dirac mode at $\mu=\mu_{c}$. The Dirac mode arises at the topological phase transition and dominates the low energy spectrum of the system, being also responsible for the universal slow dynamics\,\cite{Vodola2014,Dutta2017}. 

 Instead, for dominant hopping term $\alpha<\beta$, the critical Dirac mode is not relevant in the low energy spectrum, since the pairing term is not the leading operator in the zero momentum limit, and the equilibrium low energy theory at the critical point does not show any trace of the topological order found in the broken phase ($w=1$). However, the subleading pairing term turns out to be dangerously irrelevant and signatures of the topological order are found in the dynamics, which is always governed by the sub-leading pairing term, which is responsible for the topological transition. It is worth noting that the discrepancy between the traditional scaling argument and the \emph{anomalous universal scaling} is not related to the inapplicability of the adiabatic perturbation theory expression\,\eqref{ad_dens_exc}, as it may occur in Bose systems due to diverging occupations\,\cite{Polkovnikov2008,Bachmann2017,Defenu2018}. Rather, the \emph{anomalous universal scaling} is the consequence of deviations from the universal scaling hypotheses, see Eq.\,\eqref{scaling_forms}, occurring in LR systems. Similar deviations were already noticed in LR classical systems\,\cite{Flores-Sola2015,Flores-Sola2016}, but their consequences appear to be much more striking in the dynamics of quantum systems. 

Our results can be straightforwardly generalised to higher dimensional cases. Moreover, we expect them to be generally valid for most of the interacting $p$-wave Hamiltonians\,\cite{Sau2010,Jason2010,Lutchyn2010}, which reduce to the quadratic form of Eq.\,\eqref{h_klr} in the Bogolyubov approximation. These investigations are of fundamental importance in current technological applications, since slow dynamical manipulations of the Hamiltonian are necessary to realise MZMs\,\cite{Perfetto2013}. Finally, due to the possibility of experimentally measuring both the equilibrium and the dynamical critical scaling, the \emph{anomalous universal scaling} can be used as a diagnostic for the existence of long-range tails in the hopping matrix and of topological excitations in superconducting systems.
 \\\textit{Acknowledgements.} 
   ND is grateful to S.~Ruffo, A.~Trombettoni and G.~Gori for useful discussions at the early stages of this work.  
ND and TE acknowledge financial support by Deutsche
Forschungsgemeinschaft (DFG) via Collaborative Research Centre SFB 1225 (ISOQUANT) and under Germany’s Excellence Strategy EXC-2181/1-390900948 (Heidelberg STRUCTURES Excellence Cluster). LD acknowledges financial support from the BIRD2016 project of the University of Padova.  GM is grateful for financial support by the DFG Priority Program no.~1929 GiRyd and by the German Ministry of Education and Research (BMBF) via the Quantera project ``NAQUAS''.  Project NAQUAS has received funding from the QuantERA ERA-NET Cofund in Quantum Technologies implemented within the European Union's Horizon 2020 Programme.
\appendix
\onecolumngrid
\section{Bogolyubov Transformation}
\label{bg_trans}
The Bogolyubov transformation which diagonalises the Kitaev hamiltonian is described here in details. Our starting point is the real space Hamiltonian for $N$ spinless fermions, Eq.\,\eqref{h_klr} of the main text, which reads
\begin{align}
\label{h_klr}
H=-\sum_i\sum_{r>0}\left(j_{r}c^{\dagger}_{i}c_{i+r}+\Delta_{r}c^{\dagger}_{i}c^{\dagger}_{i+r}+{\rm H.c.}\right)-\mu\sum_{i}\left(c^{\dagger}_{i}c_{i}-\frac{1}{2}\right)\,,
\end{align}
where the $c^{\dagger}_{i}$s are Fermionic creation operators which fulfil the anticommutation relations $\{c_j,c_\ell\}=\delta_{j\ell}$. We consider power law couplings for the hopping and pairing terms:
\begin{align}
j_{r}^{\alpha}&=\frac{J}{N_{\alpha}}\,\frac{1}{\bar r^{\alpha}}\,,\label{jr}\\
\Delta_{r}^{\beta}&=\frac{d}{N_{\beta}}\,\frac{1}{\bar r^{\beta}}\,,\label{dr}
\end{align}
where $\bar r=\min(r,N-r)$ and we have considered periodic boundary conditions. The exponents of the power laws can be different and take values $\alpha>1$ and $\beta>1$, which warrant a well defined ferromagnetic state energy. The normalisation coefficients $N_\gamma$ ($\gamma=\alpha,\beta$) garantee that the energy is extensive. They read
\begin{align}
N_{\gamma}=2\sum_{r=1}^{N/2}\frac{1}{r^{\gamma}}\to2\zeta(\gamma)\,,
\end{align}
where the expression on the right is exact in the thermodynamic limit and $\zeta(\gamma)$ is the Riemann $\zeta$-function \cite{Abramowitz1964}.

Hamiltonian \eqref{h_klr} is quadratic and can be explicitly integrated in momentum space via a Bogolyubov transformation\,\cite{Vodola2014}. For this purpose we introduce the Fourier Space transformations of operators $c_j$, 
\begin{align}
\label{f_trans}
c_{k}=\frac{1}{\sqrt{N}}e^{-i\frac{\pi}{4}}\sum_{j\in\mathbb{Z}}c_{j}e^{ikj}
\end{align}
where $k\in[-\pi,\pi)$ and it takes continous values in the thermodynamic limit. Using the Fourier representation\,\eqref{f_trans} in the Kitaev hamiltonian one finds
\begin{align}
\label{h_klr}
H=\sum_k\left[(c^{\dagger}_{k}c_{k}
-c_{-k}c^{\dagger}_{-k})\varepsilon_\alpha(k)+(c^{\dagger}_{k}c^{\dagger}_{-k}+c_{-k}c_{k})\Delta_\beta(k)\right]\,,
\end{align}
where the coefficients are a function of $k$ and read:
\begin{align}
\varepsilon_{\alpha}(k)&=-\frac{\mu}{2}-j_{\alpha}(k)\,,\nonumber\\
j_{\alpha}(k)&=\sum_{r>0}j^{\alpha}_{r}\cos(kr)\,,\nonumber\\
\Delta_{\beta}(k)&=\sum_{r>0}\Delta^{\beta}_{r}\sin(kr)\,.\nonumber
\end{align}

It is convenient to employ a Bogolyubov transformation in order to diagonalise the static Hamiltonian. We choose
\begin{align}
c_{k}=u_{k}\gamma_{k}+v^{*}_{-k}\gamma^{\dagger}_{-k}\,,
\end{align}
where $u_k$, $v_k$ are the Bogolyubov coefficients and $\gamma_k$ satisfy the fermionic anticommutation relations $\{\gamma_k,\gamma_k'^\dagger\}=\delta_{k,k'}$.
With this transformation the Hamiltonian takes the diagonal form
\begin{align}
H=2\sum_{k}\omega_{k}\left(\gamma^{\dagger}_{k}\gamma_{k}-\frac{1}{2}\right)\,,
\end{align}
where the eigenfrequencies read
\begin{align}
\label{spec_eq}
\omega_{k}=\sqrt{\varepsilon_{\alpha}(k)^{2}+\Delta_{\beta}(k)^{2}}\,.
\end{align}
The diagonal form is found with the Bogolyubov coefficients
\begin{align}
(u_{k},v_{k})=\left(\cos\frac{\theta_{k}}{2}, \sin\frac{\theta_{k}}{2}\right)\,,
\end{align}
such that
\begin{align}
\tan\theta_{k}=\frac{\Delta_{\beta}(k)}{\varepsilon_{\alpha}(k)}\,.
\end{align}
This is the solution of the equilibrium model.
%
\subsection{Taylor Expansion of the Polylogarithm}
 At lowest order in $k$ (namely, for $|k|\ll \pi$) , we expand the $k$-dependent coefficients and obtain the expressions:
\begin{align}
j_{\alpha}(k)/J&=1+\sin(\alpha\pi/2)\frac{\Gamma(1-\alpha)}{\zeta(\alpha)}k^{\alpha-1}-\frac{\zeta(\alpha-2)}{2\zeta(\alpha)}k^{2}+O(k^{3})\quad\mathrm{if}\,\,\alpha<3,\label{j_exp1}\\
j_{\alpha}(k)/J&=1+\frac{2\log(k)-3}{4\zeta(3)}k^{2}+O(k^{3})\quad\mathrm{if}\,\,\alpha=3,\label{j_exp2}\\
j_{\alpha}(k)/J&=1-\frac{\zeta(\alpha-2)}{2\zeta(\alpha)}k^{2}+O(k^{\alpha-1})\quad\mathrm{if}\,\,\alpha>3,\label{j_exp3}
\end{align}
and
\begin{align}
\Delta_{\beta}(k)/d&=\cos(\beta\pi/2)\frac{\Gamma(1-\beta)}{\zeta(\beta)}k^{\beta-1}+\frac{\zeta(\beta-1)}{\zeta(\beta)}k +O(k^{3})\quad\mathrm{if}\,\,\beta<2,\label{d_exp1}\,.\\
\Delta_{\beta}(k)/d&=\frac{6(1-\log(k))}{\pi^{2}}k +O(k^{3})\quad\mathrm{if}\,\,\beta=2,\label{d_exp2}\,.\\
\Delta_{\beta}(k)/d&=\frac{\zeta(\beta-1)}{\zeta(\beta)}k+O(k^{\beta-1})\quad\mathrm{if}\,\,\beta>2,\label{d_exp3}\,.
\end{align}
 These expressions are valid for all exponents $\alpha>1$, once the analytic continuation of the Reimann $\zeta$-function is considered for the cases $\alpha\leq 2$ and $\beta<2$.  For $\beta>2$ and $\alpha>3$ the non analytic terms in Eqs.\,\eqref{j_exp1} and\,\eqref{d_exp1} become sub-leading with respect to further analytic corrections and they can safely be discarded. Now we have all the necessary information to derive a full phase diagram for the extended Kitaev chain.

\section{The scaling of the defect density} 
\label{defect_sc}
According to the solution of the effective LZ problem the excitation probability of each low momentum mode for an infinitely slow ramp is 
\begin{align}
p_{k}\approx e^{-\frac{\pi}{\delta^{2}}\Delta_{\beta}(k)^{2}}
\end{align}
and the defect density can be computed integrating the excitation probability along $k$ 
\begin{align}
n_{\mathrm{exc}}\approx\int e^{-\frac{\pi}{\delta^{2}}\Delta_{\beta}(k)^{2}}
\end{align}
in the infinitely slow ramp limit $\delta\to 0$ the above integral has to be computed using the saddle point method. Indeed, the integral remains not negligible only on an infinitesimal neighborhood of the saddle point $k=0$, where the pairing term $\Delta_{\beta}(k)$ vanishes.
According to the low momentum expansions reported in the above section for $\beta>2$ one has
\begin{align}
n_{\mathrm{exc}}\approx\int e^{-\frac{\pi}{\delta}\frac{\zeta(\beta-1)^{2}}{\zeta(\beta)^{2}}k^{2}}\approx \frac{\zeta(\beta)}{\zeta(\beta-1)}\sqrt{\delta}\propto \sqrt{\delta}
\end{align}
as it shall be for a short-range system. In the long-range regime $\beta<2$ the saddle point approximation is less straightforward due to the divergence of the Hessian in the exponent. Considering the low momentum expansion in this regime the integration reads
\begin{align}
n_{\mathrm{exc}}\approx\int e^{-\frac{\pi}{\delta}\left(\cos\left(\frac{\beta\pi}{2}\right)\frac{\Gamma(1-\beta)}{\zeta(\beta)}\right)^{2}k^{2(\beta-1)}}.\end{align}
It is convenient to define $\theta=2(\beta-1)^{-1}$ and $c=\pi\left(\cos\left(\frac{\beta\pi}{2}\right)\frac{\Gamma(1-\beta)}{\zeta(\beta)}\right)^{2}$, then we shall consider the transformation
\begin{align}
k&=(\delta\,s/c)^{\theta}\\
dk&=\delta^{\theta}\frac{\theta}{c^{\theta}}
s^{{\theta}-1}
\end{align}
the integral then reduces to
\begin{align}
n_{\mathrm{exc}}\approx\int e^{-\frac{c}{\delta}\,k^{1/\theta}}=\frac{\theta\delta^{\theta}}{c^{\theta}}\int s^{\theta-1}e^{-s}ds=\frac{\theta\Gamma(\theta)}{c^{\theta}}\delta^{\theta}\propto\delta^{\theta}\end{align}
as argued in the main text. 

At $\beta=2$ the low energy behavior for $\Delta_{\beta}(k)$ acquires logarithmic corrections and it shall then be treated separately. The low momentum limit in this case reads
\begin{align}
\label{b2_exp}
\Delta_{2}(k)=-\frac{6}{\pi^{2}}k\log(k)+O(k^{2})
\end{align}
leading to the excitation probability
\begin{align}
n_{\mathrm{exc}}\approx\int e^{-\frac{36}{\delta\pi^{3}}k^{2}\log(k^{2})}.\end{align}
In this case one shall introduce a more complicated transformation
\begin{align}
\label{log_trans}
s&=k\log(k)\\
k&=e^{W(s)}\\
dk&=\frac{ds}{1+\log(k)}=\frac{ds}{1+W(s)}
\end{align}
where $W(s)$ is the Lambert function. The integral has now been reduced to 
\begin{align}
n_{\mathrm{exc}}\approx\int\frac{e^{-\frac{36}{\pi^{3}}\frac{s^{2}}{\delta}}ds}{1+W(s)}.\end{align}
The integration boundary has to be treated with care since the transformation is not univocal. However, one shall consider that we are interested only in a small neighbourhood of $k=0$, where the expansion in Eq.\,\eqref{b2_exp} is valid. In this regime, it is sufficient to consider the lower branch of the Lambert function $W_{-1}(s)$, which is real in the interval $s\in[0,-1/e]$, leading to the momentum interval $k\in[0,1/e]$. In the $s\to 0^{-}$ limit $W_{-1}(s)$ obeys the asymptotic expansion\,\cite{Corless1996}
\begin{align}
W_{-1}(s)=\log(-s)+O(\log\log(-s))
\end{align} 
Therefore our integral can be finally approximated with
\begin{align}
n_{\mathrm{exc}}\approx\int_{0}^{-1/e} \frac{e^{-\frac{36}{\delta\pi^{3}}s^{2}}}{\log(-s)}=-\int_{0}^{1/e} \frac{e^{-\frac{36}{\delta\pi^{3}}s^{2}}}{\log(s)},
\end{align}
the reduction of the integral boundaries to $k\in[0,1/e]$ is valid for $\delta\ll\frac{1}{e^{2}}$ and becomes exact in the $\delta\to 0$ limit. In order to proceed further, it is convenient to introduce the limit representation of the logarithm
\begin{align}
\log(s)=\lim_{h\to0}\frac{s^{h}-1}{h}
\end{align}
which in turns leads to 
\begin{align}
\frac{1}{\log(s)}=\lim_{h\to 0}\sum_{n=1}^{\infty}h\,s^{hn}.
\end{align}
Once latter expression is plugged into the integral one obtains
\begin{align}
\label{b2_exc_den}
n_{\mathrm{exc}}\approx-\lim_{h\to0}\sum_{n=1}^{\infty}h\int_{0}^{1/e} s^{hn}e^{-\frac{36}{\delta\pi^{3}}s^{2}}\approx-\lim_{h\to0}\sum_{n=1}^{\infty}\int_{0}^{+\infty} s^{hn}e^{-\frac{36}{\delta\pi^{3}}s^{2}}=-\frac{\sqrt{\pi^{3}\delta}}{6}\lim_{h\to0}h\sum_{n=1}^{\infty}\left(\frac{\sqrt{\delta}}{6}\right)^{h\,n}\frac{\Gamma\left(\frac{3+h\,n}{2}\right)}{1+hn}
\end{align}
where we once again deformed the integration range, since $s\gg\sqrt{\delta}$ contributions to the integral vanish exponentially fast in the $\delta\to 0$ limit. The summation in Eq.\,\eqref{b2_exc_den} has to be considered in the $h\to 0$ limit, where the power law contributions $\delta^{hn/2}$ become all relevant, while the $\frac{\Gamma\left(\frac{3+h\,n}{2}\right)}{1+hn}$ terms can be safely approximated as $\Gamma\left(\frac{3}{2}\right)=\sqrt{\pi}/2$ yielding
\begin{align}
\label{b2_exc_den_c}
n_{\mathrm{exc}}\approx-\frac{\pi^{2}}{12}\sqrt{\delta}\lim_{h\to0}h\sum_{n=1}^{\infty}\left(\frac{\sqrt{\delta}}{6}\right)^{h\,n}=-\frac{\pi^{2}}{6}\frac{\sqrt{\delta}}{\log(\delta/6)}\propto -\sqrt{\delta}\log(\delta)^{-1}.
\end{align}
As expected the case $\beta=2$ is exactly in between the pure short-range case $n_{\mathrm{exc}}\propto \sqrt{\delta}$ and the weak long range case $\beta=2-\varepsilon$, where one has $n_{\mathrm{exc}}\propto \delta^{\frac{1+\varepsilon}{2}}$ for $\varepsilon\ll 1$ and, therefore the excitation probability decays faster than in the pure short-range case. The $\varepsilon\to0$, i.e. $\beta\to 2$, limit is placed exaclty in between, with the density of excitation decaying only logarithmically faster than in the short-range case. We have numerically verified that the introduction of sub-leading $k^{2}$ terms in the expansion of $\Delta_{2}(k)$ as well as the extension of the integration range beyond the region of validity of the low momentum expansion in Eq.\,\eqref{b2_exp} do not modify the scaling regime in the $\delta\to 0$ limit.

\section{Landau-Zener Problem}
\label{lz}
As discussed in the main text, one can employ the substitution in Eq.\,(12) into the dynamical evolution Eq.\,(11) for the Bogolyubov coefficients. The resulting dynamics takes the celebrated Landau-Zener form
\begin{align}
\label{lz_dyn}
i\partial_{\tau}\begin{pmatrix}u_{k}\\
v_{k}\end{pmatrix}=\begin{pmatrix}-\Omega_{k}\tau & 1\\
1 & \Omega_{k}\tau\end{pmatrix}\begin{pmatrix}u_{k}\\
v_{k}\end{pmatrix}.
\end{align}
The Landau-Zener (LZ) evolution, Eq.\,\eqref{lz_dyn}, can be solved exactly using several approaches\,\cite{Zener1932,Wittig2005,damski2005}. However, this exact solution is rather cumbersome, since it is obtained in terms of Weber functions. Therefore, we will rather rely on an approximate solution, which is capable to correctly reproduce the defect scaling, only sacrificing the exactness of numerical coefficients, unimportant to our scopes. The LZ Hamiltonian can be conveniently written using Pauli's matrices
\begin{align}
H_{LZ}=\Omega\tau\sigma_{z}+\sigma_{x}
\end{align}
where
\begin{align}
\sigma_{z}=\begin{pmatrix}
1 & 0\\
0 & -1
\end{pmatrix},\quad \sigma_{x}=\begin{pmatrix}
0 & 1\\
1 & 0
\end{pmatrix}.
\end{align}
This Hamiltonian is diagonalised analogously to the Bogolyubov transformation described in the first section, one has to introduce the angle $\theta=\mathrm{arctan}\left(\frac{1}{\Omega\tau}\right)$, useful to describe the eigenstates
\begin{align}
|+\rangle=\begin{pmatrix}
\cos\frac{\theta}{2}\\
\sin\frac{\theta}{2}
\end{pmatrix},\quad |-\rangle=\begin{pmatrix}
\cos\frac{\theta}{2}\\
-\sin\frac{\theta}{2}
\end{pmatrix}
\end{align}
with the eigen-energies $\omega_{\pm}=\pm\sqrt{(\Omega\tau)^{2}+1}$.
\section{The finite ramp case}
\label{fr_corr}
 At the initial stage of the dynamics the system is exactly in the ground state $|\psi_{\tau=0}\rangle\equiv |-\rangle$, while at every finite time the state is given by the superposition $|\psi_{t}\rangle = \alpha_{-}(\tau)|-\rangle+\alpha_{+}(\tau)|+\rangle$, with $ \alpha_{+}(\tau)^{2}+ \alpha_{-}(\tau)^{2}=1$. According to adiabatic perturbation theory\,\cite{de_grandi2010}, the excitation amplitude at first order in $\Omega$ is given by the formula
\begin{align}
\alpha_{+}(\tau)\simeq - \int_{\tau_i}^{\tau}\langle+|\partial_{s} |-\rangle e^{i(\Theta_{+}(s)-\Theta_{-}(s))}
\end{align}
where
\begin{align}
\Theta_{\pm}(\tau)=\int_{\tau_{i}}^{\tau}\omega_{\pm}(s)ds.
\end{align}
The overlap element between the adiabatic states can be computed exactly for the LZ model,
\begin{align}
\langle+|\partial_{\tau} |-\rangle=\frac{\partial_{\tau}\theta}{2}=-\frac{1}{2}\frac{\Omega}{(\Omega\tau)^{2}+1}
\end{align}
yielding the transition amplitude
\begin{align}
\label{trans_amp}
\alpha(\tau_{f})\simeq \frac{1}{2}\int_{-\infty}^{\tau_{f}}\frac{\Omega\,d\tau}{(\Omega\tau)^{2}+1}e^{\frac{2i}{\Omega}\int_{0}^{\Omega\tau}\sqrt{s^{2}+1}ds}.
\end{align}
where, without loss of generality, we imposed $\tau_{i}=-\infty$. One can explicitly integrate the phase factor
\begin{align}
g(x)=2\int_{0}^{x}\sqrt{s^{2}+1}ds=\tau\sqrt{\tau^{2}+1}+\mathrm{arcsinh}(\tau).
\end{align}
It is convenient to rescale the integration variable in Eq.\,\eqref{trans_amp} according to $x=\Omega\tau$,
\begin{align}
\alpha(\tau_{f})\simeq \frac{1}{2}\int_{-\infty}^{\Omega\tau_{f}}\frac{dx}{x^{2}+1}e^{\frac{i}{\Omega}g(x)},
\end{align}
we are interested in the $\tau_{f}\gg 0$ limit of the latter we shall then separate the integral into the two contributions
\begin{align}
\label{amp_2}
\alpha(\tau_{f})\simeq \frac{1}{2}\int_{-\infty}^{\infty}\frac{dx}{x^{2}+1}e^{\frac{i}{\Omega}g(x)}- \frac{1}{2}\int_{\Omega\tau_{f}}^{\infty}\frac{dx}{x^{2}+1}e^{\frac{i}{\Omega}g(x)}.
\end{align}
The above expression proves that the finite ramp dynamics is always equivalent to an infinite ramp from $\tau_{i}=-\infty$ to $\tau_{f}=+\infty$ plus a correction, which is equivalent to the exctitation amplitude of a finite ramp not crossing the critical point. The phase factor $g(x)$ has no stationary points on the real line, but it possesses an inflection point at $x=0$. Therefore, the first contribution to Eq.\,\eqref{amp_2} needs to be treated separately. This computation has been already carried on in details in Ref.\,\cite{de_grandi2010}, yielding 
\begin{align}
\label{inf_ramp}
\alpha(\infty)\simeq \frac{\pi}{3}e^{-\frac{\pi}{2\Omega}}
\end{align} 
where the numerical coefficient $\pi/3\approx 1.05$ is surprisingly close to the exact value $1$. The second contribution can be transformed into
\begin{align}
\label{finite_ramp}
\alpha^{*}(-\tau_{f})=\frac{1}{2}\int_{-\infty}^{-\Omega\tau_{f}}\frac{dx}{x^{2}+1}e^{\frac{i}{\Omega}g(x)}.
\end{align}
Since $\tau_{f}$ is positive $-\tau_{f}$ is negative and the formula describes the excitation amplitude of a ramp ending below the critical point. Therefore, the integration in Eq.\,\eqref{finite_ramp} does not contain the higher order stationary point $x=0$ and it can be safely integrated using the standard procedure for fast oscillating integrals\,\cite{Dingle1975}
\begin{align}
\label{finite_ramp_res}
\alpha^{*}(-\tau_{f})=\frac{\Omega}{4}\frac{1}{\sqrt{1+(\Omega\tau_{f})^{2}}^{3}}.
\end{align}
Coming back to the Kitaev chain problem one has $\Omega\equiv \delta/\Delta(k)^{2}$ and $\tau_{f}=(j_{\alpha}(k)-g_{f})\Delta_{\beta}(k)/\delta$. The excitation probability for a single momentum state $k$ is
\begin{align}
p_{k}=\frac{\delta^{2}}{16}\frac{\Delta_{\beta}(k)^{4}}{(\Delta_{\beta}(k)^{2}+(j_{\alpha}(k)-g_{f})^{2})^{3}}.
\end{align}
Therefore, the defect density for the $p$-wave superconducting Hamiltonian in Eq.\,\eqref{h_klr} after a quench starting at $g_{i}=+\infty$ and ending at $g_{f}>1$ without crossing any critical points is given by
\begin{align}
n_{\mathrm{exc}}(t_{f})=\int p_{k}dk=\frac{\delta^{2}}{16}\int\frac{\Delta_{\beta}(k)^{4}dk}{(\Delta_{\beta}(k)^{2}+(j_{\alpha}(k)-g_{f})^{2})^{3}}
\end{align}
where, as long as $|g_{f}|>1$ the integral remains always convergent. In the $\delta\to 0$ limit such contribution decays quadratically with $\delta$, in agreement with our expectations for adiabatic dynamics. The latter result proves that finite ramp corrections are always negligible with respect to the non analytic scaling in the defect density generated by the low momenta during the full ramp dynamics.
\twocolumngrid
\bibliographystyle{apsrev_cm}
\bibliography{KZM_LR_KC}

\end{document}